# ULTRAHIGH INTERFERENCE SPATIAL COMPRESSION OF LIGHT INSIDE THE SUBWAVELENGTH APERTURE OF A NEAR-FIELD OPTICAL PROBE

N.M.Arslanov, S.A.Moiseev

Zavoisky Physical-Technical Institute of the Russian Academy of Sciences,
Sibirsky Trakt str. 10/7, Kazan, 420029, Russian Federation
**E-mail:** narslan@mail.ru, samoi@yandex.ru

## ABSTRACT

Spatial effects of interference and interaction of light modes in the subwavelength part of the near-field optical microscopy probe have been theoretically studied. It was found that the mode interference can lead to higher spatial compression of light ($\lambda$ = 500 nm in free space) within the transverse size of 25 nm inside the probe output aperture of 100 nm in diameter. The results principally demonstrate the possibility of increasing spatial resolution of the near-field optical microscopy technique.

**Keywords:** near-field optics, light interference effects, ultrahigh spatial compression of light.

**PACS numbers:** 07.79.Fc; 87.64xx; 42.25 –p.

## 1. INTRODUCTION

The near-field optical (NFO-) microscopy technique attracts a great attention in experimental investigations of the physical properties of material surfaces [1, 2, 3]. In a most well-known variant of NFO-microscopy technique the light propagates through the optical fiber to the probe with taper tip coated with metal. Diameter of the probe output aperture is significantly smaller than the wavelength $\lambda$, therefore, the light may be localized on the material surface within the area $S \ll \lambda^2$. The narrowing part of the NFO-microscopy probe has transverse sizes much smaller than the cut-off radius of the majority modes [3] propagating in such optical fiber, thus only a small part of the light energy gets into the probe output aperture. The low energy throughput is the main factor limiting the spatial resolution of the technique and worsening the polarization characteristics of the output radiation. At present it is well known that presence of the evanescent field enables such spatial light localization in the probe. Experimental measurements of the output far-field parameters [4, 5, 6] show that the evanescent light acquires unusual spatial characteristics in the probe aperture [3].

However, it is also well-known that the field scattering diagram has a spread diffuse character at the output aperture. Experimental investigations of the spatial structure of light are considerably restricted in the near-field zone of the subwavelength probe aperture. In addition we note that the influence of the spatial shape and physical parameters of the probe on the propagation and spatial structure of the light in the probe output aperture have still been insufficiently studied in spite of the theoretical works [7, 8, 9, 10, 11, 12]. Therefore, clear physical understanding of the spatial properties of the compressed light in the probe is still absent which significantly restricts development of the (NFO-) microscopy methods. In this paper we theoretically study the interaction and interference effects of the light modes in the subwavelength probe. In the first part of the paper an analytical method is developed on the basis of the B.Z.Katzenellenbaum's *cross-sections* theory [13]. Using the developed theoretical approach we make a numerical analysis of the influence of the light interference effects on the spatial structure of the light in the probes with different geometry. Our results demonstrate that the light interference effects may play an important role in an sharper spatial compression of the light inside the subwavelength probe aperture.

## 2. BASIC EQUATION OF THE CROSS-SECTION METHOD

The B.Z. Katzenellenbaum's method [13] enables one to study light propagation in the probe taking into account its real spatial and physical properties [14, 15]. In this method the light field in the arbitrary coordinate z of the narrowing probe is represented through the superposition of the light waves propagating in both (z, -z) directions:

$$\vec{E}(z) = \sum_{-\infty}^{\infty} P_j(z)\vec{E}^j(z), \quad \vec{H}(z) = \sum_{-\infty}^{\infty} P_j(z)\vec{H}^j(z), \tag{1}$$

where the field components $\vec{E}^j(z)$ and $\vec{H}^j(z)$ correspond to the j-th mode of the auxiliary comparison waveguide, which has the same spatial distribution in the probe cross-section at the point z.

The total fields E(z) and H(z) satisfy the same boundary conditions in a certain surrounding of the tilted wall as the field modes $E^j$ and $H^j$ on the comparison waveguide wall [13]. The mode amplitudes $P_j(z)$ at the beginning (z=0) and the end (z=L) of the irregular part of the waveguide are equal to the appropriate mode amplitudes in the regular part of the probe. Spatial evolution of the amplitudes $P_j(z)$ in the probe is determined by the following system of equations:

$$\partial_z P_j(z) - ih_j(z)P_j(z) = \sum_{v=-\infty}^{v=\infty} S_{jm}(z)P_m(z), \qquad (2)$$

where $S_{jm}$ (z) is a coefficient coupling the modes on the probe wall:

$$S_{jm}(z) = \frac{a'(z)a(z)}{2h_j(z)(h_j(z)-h_m(z))} \oint_C d\varphi \left(1-\frac{\varepsilon_o}{\varepsilon}\right)(H_z^j H_z^m - H_\varphi^j H_\varphi^m + \varepsilon_0 E_r^j E_r^m) \cdot \qquad (3)$$

Here we have taken into account that the NFO-microscopy probes can be considered as nonmagnetic material with $\mu_0 = \mu = 1$ (where $\mu_0, \mu$ are the magnetic permeabilities of the probe core and the metallic coating). The probe narrowing is chosen uniform over the probe cross section; r, φ, z - are cylindrical coordinates; a(z) is a waveguide radius in z-coordinate, a`(z) is a corner tangent of the probe wall to the longitudinal axis z; $h_j(z)$ is a wavenumber of the mode j-th, which has the following analytical solution for the waveguide with an ideal metal wall:

$$h_j(z) = \sqrt{k_o^2 \varepsilon_o - \alpha_j^2(z)}, \qquad (4)$$

where $k_0 = \omega / c$ is a wavenumber of the light; $\varepsilon_o$ and ε are the dielectric permeabilities of the probe core and metallic coating. We assume that $\varepsilon_o$=2.16 and ε=-34.5+8i, which corresponds to the aluminum coating of the probe with thickness more than a skin-layer typically equal to 6 nm [16]; $\alpha_j(z) = \nu_j / a(z)$ is the own number of $TM_j$ mode in the probe with the ideal wall (index j denotes two parameters (n,l): n= 0, 1, , l = 0,1, ), where $\nu_{j=(n,l)}$ is *l*-th root of the n-th order Bessel function $J_n(x)_{x=\nu_j} = 0$; $\alpha_j(z) = \mu_j / a(z)$ is the own number of $TE_j$ mode, $\mu_{j=(n,l)}$ is the j-th root of the derivative of n-th order Bessel function corresponding to the $TE_j$ mode ($\partial/\partial x\, J_n(x)_{x=\mu_j} = 0$).

Unlike regular waveguide, our calculations have shown that the rapid change of the boundary conditions in the NFO-probe causes effective energy redistribution of the light into several coupled light field modes, which interference results in the spatial compression of the light. The system of equations (2) describes the interaction between the forward spatial modes and the backward modes, which are reflected from the narrowing walls of the probe. Each light mode interacts with all forward and backward modes of the HE- or EH- types as well as each mode is absorbed in the probe walls. Description of the field modes evolution according to Eqs. (2) in the probe requires calculation of the values $S_{jm}$ and $h_j$ taking into account real physical parameters. We can calculate the coupling coefficient $S_{jm}$ using the fact that the magnetic part of the electric modes has the second order on wall wave resistance $\xi$ ($\xi = \sqrt{\mu/\varepsilon} <<1$) in the Taylor's decomposition on $\xi^n$ [13, 17]. Similarly, the electric field component of the magnetic modes represents the second order magnitude on the wall wave resistance $\xi$.

Therefore in order to find the expression for the coupling coefficient in the first order of $\xi$ it is possible to use the modes of the ideal waveguide [16] and calculate the wavenumber of the modes in the first order of $\xi$. Thereby the coupling coefficients of the field modes can be found using solutions for the wavenumbers of the waveguide modes with the wall made of non-ideal metal. In a general case the wavenumbers are the roots of the complex transcendental equation obtained from the boundary conditions for the field modes on the metallic wall of the regular waveguide. The equation was solved numerically in [**19**]. We use Leontovich boundary condition [**18**] in order to determine the mode wavenumbers $h_j$ in the waveguide with non-ideal walls [13]. In the cylindrical coordinate system the condition has the following form of the series up to the first-order of $\xi$:

$$E_\varphi = \xi H_z, \; E_z = -\xi H_\varphi . \tag{5}$$

Writing the expressions for the field modes in Eq.(5) through the well known Hertz potentials and using the Green integral (see Appendix A) we find the following expression for the wavenumber of TM modes:

$$h_j(z) = \sqrt{k_o^2 \varepsilon_o - v_j^2 / a(z)^2 + \xi 2ik_0 \varepsilon_0 / a(z)}, \tag{6}$$

and for the TE modes:

$$h_j(z) = \sqrt{k_o^2 \varepsilon_o - \mu_j^2 / a(z)^2 + \xi \frac{2a(z)i}{k_0(\mu_{nj}^2 - n^2)} \left( \frac{\mu_{nj}^4}{a(z)^4} + h_0^2(z) \frac{n^2}{a(z)^2} \right)}, \tag{7}$$

The obtained analytical solutions (6), (7) are demonstrated on the Fig.1 and Fig.2, which with high accuracy coincide with the earlier numerical results of the work [**19**]. The solutions (6) and (7) allow us to significantly facilitate the numerical analysis of the field evolution in the near-field microscopy probe, studied below for a number of specific probe forms.

## 3.SPATIAL INTERFERENCE OF LIGHT MODES IN THE PROBE

The total field energy decreases when the field propagates through the probe towards the output aperture. According to the Eq. (1) the field intensity in the arbitrary point with the coordinate z, $r_\perp$ is given by the expression:

$$I(\vec{r}) = \sum_j |P_j(z)|^2 \vec{E}_j^2(\vec{r}) + \{\sum_{j<j'} P_j(z)P_{j'}^*(z)\vec{E}_j(\vec{r})\vec{E}_{j'}^*(\vec{r}) + \kappa.c.\}. \quad (8)$$

Below we study the situation where the main mode $P_j=\delta_{j,01}$ is excited at the probe entrance.

The excited main mode has the least decay in the fiber among the modes propagating in the probe. In accordance with the calculation given below, the field propagation ( $\lambda$ = 500 nm. in free space) starts to decrease rapidly after the probe cross-section radius reaches the critical value for the main mode $a(z_{\kappa p}) = v_1\lambda/(2\pi\sqrt{\varepsilon_o})$ , (below $a(z_{\kappa p}) \approx \lambda/4$ ). The comparison of the coupling coefficients of the field modes (3) gives that $S_{j,j\pm1}(z) >> S_{j,j\pm2}(z)$, thus the interaction between the nearest modes plays a crucial role in the dynamics of the field propagation in the near-field optical probe. Particularly strong influence on the field mode propagation occurs in the small output domain of the probe where the mode wavenumbers become complex values and the most significant decay occurs for the modes. At the same time as it follows from the analysis of the equations system (2), the energy exchange between the field modes exceeds the decay of each mode in the probe. Therefore it is reasonable to estimate the possibility of efficient generation of the nearby light modes at the field propagation in the probe and analyze the appearance of the multimode light structure in the probe output.

It is well-known that the light throughput increases, as the inclination of the probe wall does [**3**, **20**, **21**, **22**]. Herewith in our opinion the probe shape variation should have a great impact on the field mode structure in the probe. Below in the Fig.3 we demonstrate the numerical calculation of the power $\vec{S}(z,\lambda) = \frac{c}{8\pi}\mathrm{Re}\iint_\Sigma [\vec{E}\vec{H}^*]d\vec{S}$ of the excited $TM_{0m}$ modes for the different inclination of the probe wall. The calculations have shown that the main light energy is carried in the probe by the three lowest field modes due to the small values of the coupling coefficients $S_{1j}$ $S_{2j}$ $S_{3j}$ (j>3) comparing to the coefficients $S_{12}$ , $S_{13}$ and $S_{23}$. Increasing the probe wall inclination is accompanied by the growth of the coupling coefficients that causes a considerable equalization of the energy between the three main modes at the probe output. The second field mode becomes comparable with the first main mode when the angle value α is close to 75° - 76° as shown in the Fig. 3. We shall note that the second light mode is excited more intensively than the third mode noticeably increasing only for angle α larger than 72°. Herewith the interaction of the third mode with the second mode fails to strongly influence upon the second mode energy. Under such condition, using the results of work [23] it is possible to show that the influence of the third mode on the spatial structure of light field in the probe output is also strongly suppressed by considerable reflection of the mode from the output aperture. Accordingly, it is possible to ignore the influence of the third mode on the light spatial structure of the output radiation (in Fig. 3.) for angles less than 80.

Accomplished calculations of the total energy transferred by all field modes are in accordance with available results [**3**, 20, 21, 22]. Our calculation of the light spatial structure for the $TM_{11}$ modes are in good agreement with experimental data [5, 6] of the measurement of the far-field intensities. We have studied the interference peculiarities of the light modes in the probe and its influence upon the light field spatial structure at the probe output with the output aperture diameter $D=2a(z=L)=$ 100 nm. It turned out that the interference effects of the light modes are imperceptible for the inclination α< 45°, when the whole light energy is carried by first main mode. In this case the distinctive spatial structure of the light at the probe output aperture is shown in the Fig. 4, where the intensity of longitudinal field component $I_z(r,\varphi,z) \sim |P_1(z)|^2 J_0(\alpha_{01}r)^2$ and radial component $I_r(r,\varphi,z) \sim |P_1(z)|^2 J_1(\alpha_{01}r)^2$. (where $J_n(\alpha_{01}r)$ are Bessel functions of n-th order, which describe the field $TM_{0m}$ modes in accordance with the boundary conditions [16]).

Our numerical calculations have shown that the increasing of the angle α up to 65° leads to the amplification of the second field mode. Taking into account only the two lowest light modes, we obtain the intensities of the longitudinal $I_z(r,\varphi,z)$ and radial $I_r(r,\varphi,z)$ components at the probe output for the aperture radius $a(z=L) << a_{kr} = v_j /(k_o \sqrt{\varepsilon_o}) \approx \lambda/3$:

$$I_z(r,\varphi,z) \sim |P_1(z)|^2 J_0(\alpha_{01}r)^2 + |P_2(z)|^2 J_0(\alpha_{02}r)^2 + 2|P_1(z)P_2(z)| J_0(\alpha_{01}r) J_0(\alpha_{02}r)\cos(\theta_1-\theta_2), \quad (9)$$

$$I_r(r,\varphi,z) \sim |P_1(z)|^2 J_1(\alpha_{01}r)^2 + |P_2(z)|^2 J_1(\alpha_{02}r)^2 + 2|P_1(z)P_2(z)| J_1(\alpha_{01}r) J_1(\alpha_{02}r)\cos(\theta_1-\theta_2), \quad (10)$$

where $\theta_i$ is a phase of i-th mode.

However, the transverse size of the light field (refer to. the Fig. 5) is still defined by spatial structure of the longitudinal component of the main mode with the thickness about d ≈ 80 nm on half intensity. New pattern of the light field behavior arises for larger inclinations of the probe wall: $65^o <$ $\alpha< 80^o$, when the second mode energy significantly increases. Behaviour of the mode amplitudes $|P_j(L)|^2$ is shown in the Fig. 6 depending on α. Interference terms in Eq. (9) and Eq. (10) are proportional to $2|P_1(L)P_2(L)\cos\{\theta_1(L)-\theta_2(L)\}|$ and essentially influence the field characteristics in the probe output.

The numerical calculations have shown that the phase difference between the two field modes is $\theta_1(L)-\theta_2(L)\approx\pi$ for the angle $\alpha< 76^o$ (see Fig. 7.) in accordance with the relation of the coupling coefficients between the modes: $S_{12}(z)=-S_{21}(z)(h_2(z)/h_1(z))\big|_{a(z)<<a_{kr}}\sim -S_{21}(z)$, which follows from Eq. (2) and Eq. (3). Due to such phase relation and comparability of the mode amplitudes $P_1(L)$ and $P_2(L)$, their interference considerably influences the light field spatial structure. The interference leads to a sharp *spatial compression* of the light field for the angle $\alpha\approx 75^o$ (see Fig. 8).

The additional consequence of the interference is that the main part of the light field at probe the output aperture is principally determined by longitudinal field component, which aquires a spatial toric structure with *ring* thickness d ≈ 25 nm. This *ring* thickness determines spatial limits of the light compression, which can be noticeably narrower than the output probe diameter D=100 nm. The spatial thickness of the longitudinal $d_z(z)$ and transverse $d_r(z)$ components of the field intensities on its half height are shown in Fig. 9 and Fig.10. As seen from Fig. 9 appearance of the spatial torus is the result of the two light modes interference, which dramatically changes the spatial width $d_z(z)$ at the angle $\alpha\approx 55^o$. Therewith it seen from Fig. 10, Fig. 4, Fig. 5 and Fig.8 that the value of $d_r(z)$ decreases monotonous while the main energy of the transverse polarized field component is restricted by the probe walls.

We note the main intensity peak would be only in the probe output centre with the spatial width about 80 nm (for D=100 nm) for other phase relation between the field modes (particularly for the phases $\theta_1(L)-\theta_2(L)\approx 0$). This behaviour would be also accompanied by additional weaker intensity peaks near the aperture boundary. Analytical and numerical analysis of the light properties becomes too complicated for the angle $\alpha > 85^{o}$ due to the reflection effects of the field modes from the probe output. We will study these effects elsewhere later.

## 4. CONCLUSION

In this work the affects of interaction and interference of the light modes on the spatial structure of the output field in near-field optical microscopy probe have been studied. The presented investigation have been performed for the real physical and spatial parameters of the probe with diameter of the output aperture D=100 nm, which is noticeably smaller than the light wavelength $\lambda$=500 nm. The investigation has shown that the change of the probe shape (increasing of the wall inclination $\alpha$ near the probe output aperture) considerably enhances the interaction and interference of the spatial light modes in the narrowing part of the probe, which in turn dramatically changes the spatial structure of the light field. Such effects already reveal itself via the presence of two mode structure of the light field at the wall inclinations $\alpha > 45^{o}$. The two mode interference effects are especially amplified for the angels $65^{o} < \alpha < 80^{o}$ leading to the narrow spatial toric structure in the field intensity on the probe output. The toric structure demonstrates a stronger spatial compression of the light for the angle $\alpha = 75^{o}$ where the toric wall has minimum thickness close to 25 nm. Such spatially compressed light contains areas with primarily longitudinal or transverse polarization. We note that spatial compression of the field from the diameter $D_1$=100 nm up to $D_2$=25 nm will decrease the light throughput to $\sim (D_1 / D_2)^2 = 256$ times [3, 7] if we use a usual direct reduction of the probe output aperture.

Modern experimental methods offer good opportunities to control distance $\delta z$ with accuracy of nanometers between the medium surface and the plane of the output probe aperture [1]. For sufficiently small distances $\delta z \ll \lambda$, the light field on the surface will coincide practically with the light field at the output probe aperture. Thus by choosing the proposed near field probe with the inclination $\alpha \approx 76^{o}$ one can considerably enhance the spatial resolution in the experiments with small objects (like single molecules and other nanoparticles) where the resolution will be determined by the narrow peaks of the field intensities $d_z(\alpha)$. The predicted ultrahigh "interference" spatial compression of the light field offers a new attractive possibility for the experimental investigation of solid states surfaces with nanometer resolution. Detailed study of these problems represents an interesting subject for further investigations.

This work was supported by the Russian Foundation for Basic Research grants No: 03-03-96214 and grant NIOKR Tatarstan №06-6.3-343/ 2005Ф (06).

## APPENDIX A: WAVENUMBER IN PROBE WITH NON-IDEAL METAL WALL

Here we present the field components on the boundary using Hertz potentials [16] and find the condition, which couples these functions:

$$ih\frac{\partial\Pi^e}{r\partial\varphi}-ik_0\mu_0\frac{\partial\Pi^m}{\partial r_1}=\xi\alpha^2\Pi^m, \qquad (A.1)$$

$$\alpha^2\Pi^e=-\xi\left(ik_0\varepsilon_0\frac{\partial\Pi^e}{\partial r}+ih\frac{\partial\Pi^m}{r\partial\varphi}\right). \qquad (A.2)$$

Hertz potentials satisfy the equation:

$$\Delta_\perp\Pi_n+\alpha_n^2\Pi_n=0 \qquad (A.3)$$

and normalization condition:

$$\alpha_n^2\int dS\cdot\Pi_n{}^{TM}\Pi_n{}^{TM}=1,\ \alpha_n^2\int dS\cdot\Pi_n{}^{TE}\Pi_n{}^{TE}=\varepsilon_0, \qquad (A.4)$$

where $\alpha_n^2=k_0^2\varepsilon_0-h_n^2$ is the own numbers of $TM_j$ modes.

Auxiliary function $\Phi^e=h\Pi^e$ is introduced in order to determine the wavenumber of magnetic waves. Therefore we obtain the following system of equations:

$$i\frac{\partial\Phi^e}{r\partial\varphi}-ik_0\frac{\partial\Pi^m}{\partial r}=\xi\alpha^2\Pi^m, \qquad (A.5)$$

$$\alpha^2\Phi^e=-\xi\left(ik_0\varepsilon_0\frac{\partial\Phi^e}{\partial r}+ih^2\frac{\partial\Pi^m}{r\partial\varphi}\right), \qquad (A.6)$$

$$\Delta\Phi^e+\alpha^2\Phi^e=0, \qquad (A.7)$$

$$\Delta\Pi^m+\alpha^2\Pi^m=0. \qquad (A.8)$$

Expanding the own numbers and functions Φ and Π into the Taylor series on $\xi \ll 1$

$$\alpha^2=\alpha_0^2+\xi\alpha_1^2+\dots,\quad h^2=k_0^2\varepsilon_0-\alpha^2=h_0^2-\xi\alpha_1^2-\dots. \qquad (A.9)$$

$$\Phi=\Phi_0^e+\xi\Phi_1^e+\dots,\qquad \Pi^m=\Pi_0^m+\xi\Pi_1^m+\dots, \qquad (A.10)$$

and solving the system of Egs. (21)-(24) using Green formula $\oint\!\!\oint_S dS\left(V\Delta U - U\Delta V\right) = \oint_C ds\left(V\frac{\partial U}{\partial n} - U\frac{\partial V}{\partial n}\right)$ we find the following expression for parameter $\alpha_1^2$:

$$\alpha_1^2 = \frac{i}{k_0\mu_0}\left(-\alpha_0^4\oint_C rd\varphi\cdot\Pi_0^{m2} + h_0^2\oint_C rd\varphi\cdot\Pi_0^m\frac{\partial^2\Pi_0^m}{r^2\partial\varphi^2}\right) = \frac{-2ai}{k_0(\mu_{nj}^2 - n^2)\cdot}\left(\alpha_0^4 + h_0^2\frac{n^2}{a^2}\right). \qquad \text{(A.11)}$$

Performing similar calculations for electric waves we obtain the following result:

$$\alpha_1^2 = -ik_0\varepsilon_0\oint_C rd\varphi\left(\frac{\partial\Pi_0^e}{\partial r}\right)^2 = -2ik_0\varepsilon_0 / a\,. \qquad \text{(A.12)}$$

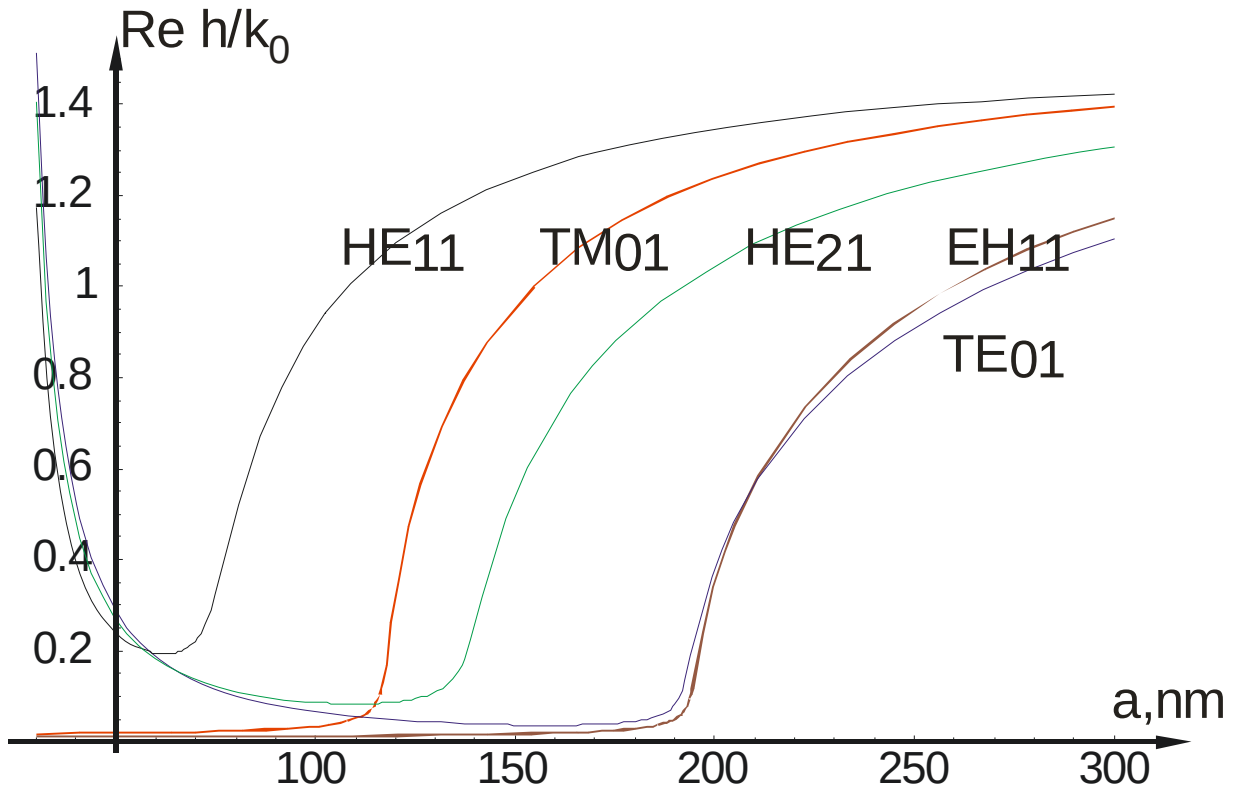

Fig. 1. Dependence of the wave number real part Re h on radius a.

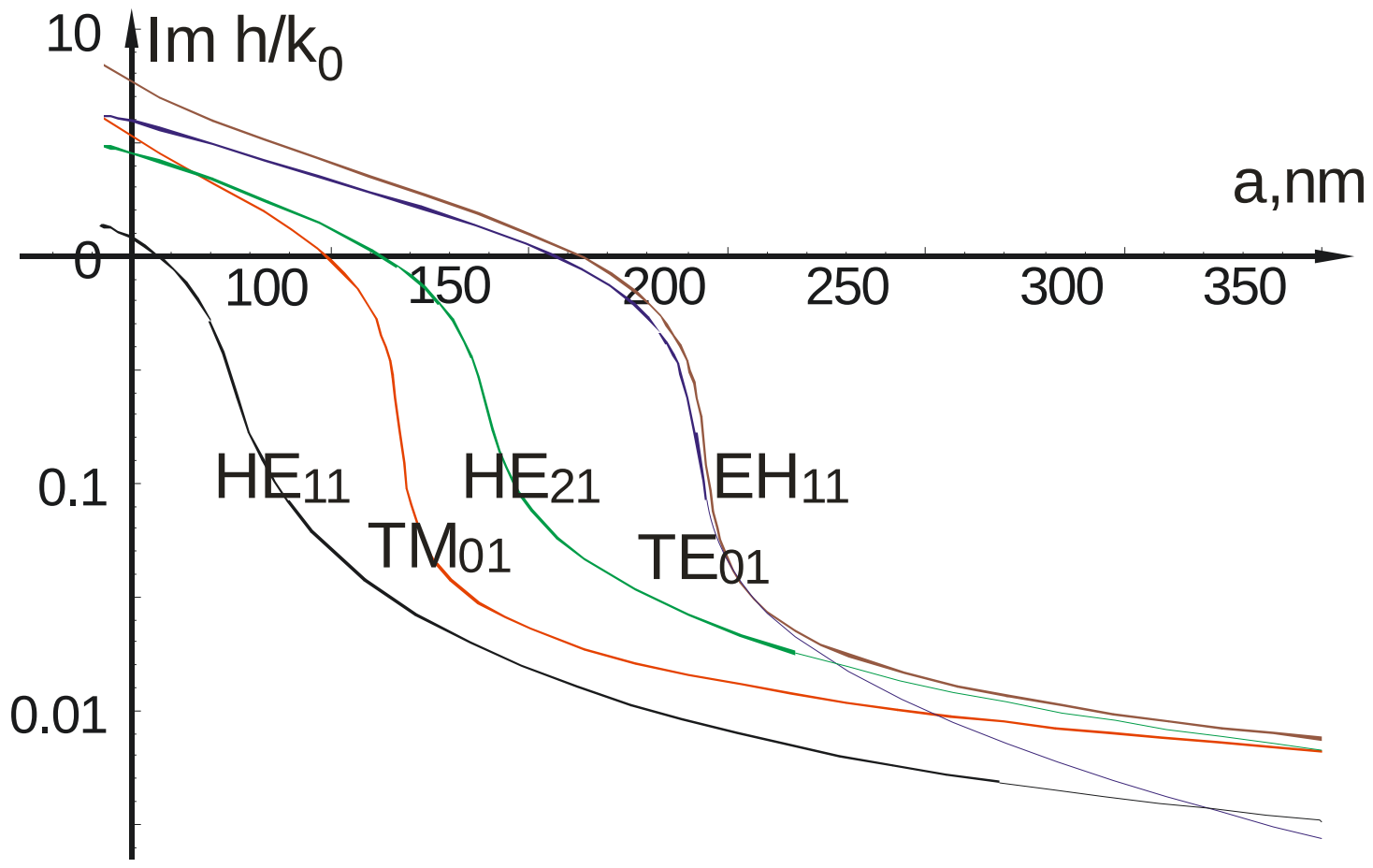

Fig. 2. Dependence of the wave number imaginary part Im h on radius a.

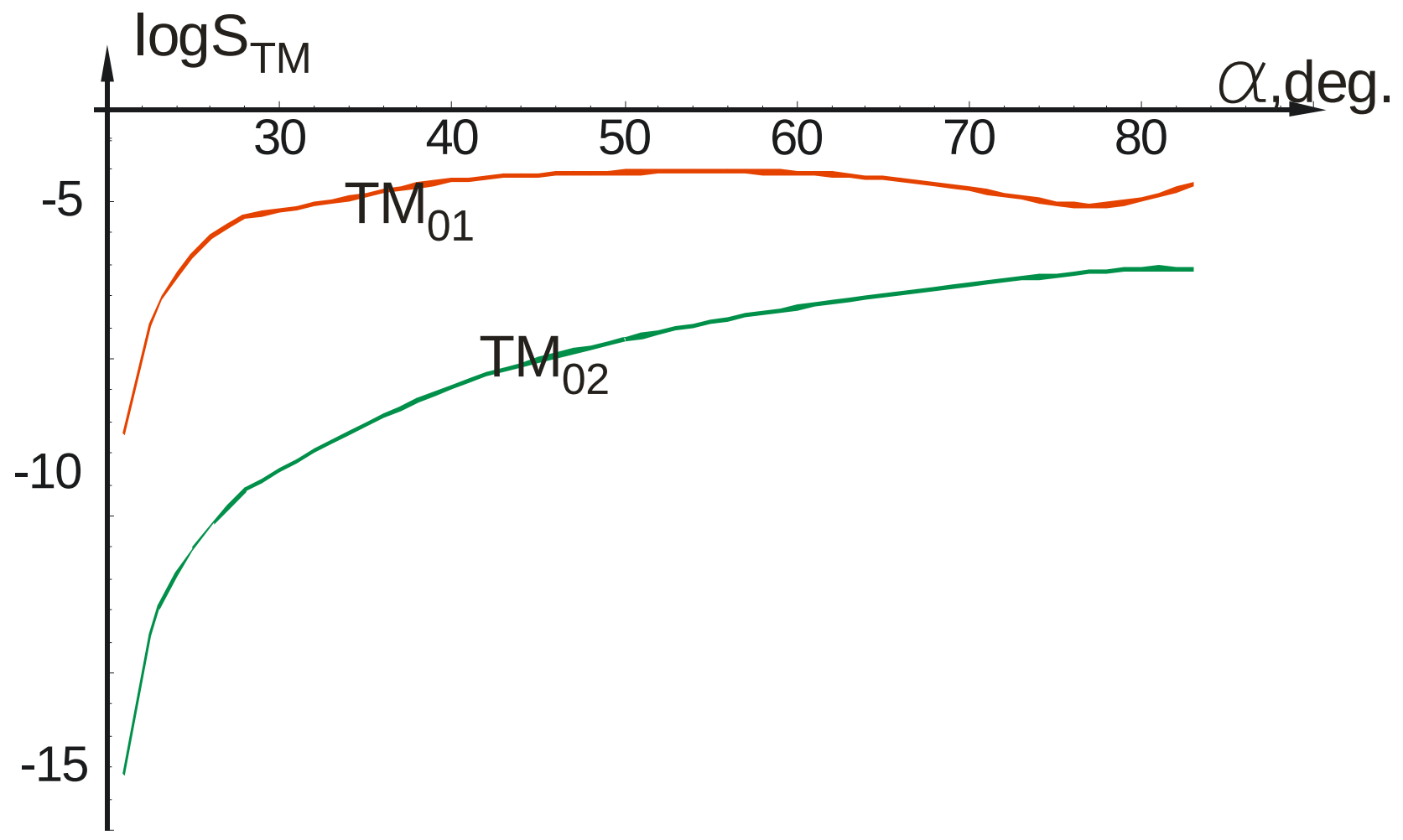

Fig. 3. Results of the numerical calculations of the field mode powers $TM_{0m}$ (m =1,2,3) (wavelength of light λ=500 nm) in the probe output with aluminum coating. Length of the probe (L) was changed from L=1172 nm to L= 8 nm with initial radius a(z=0)=500 nm and output radius a(L)=50 nm., which correspond to the change inclination α of the probe.

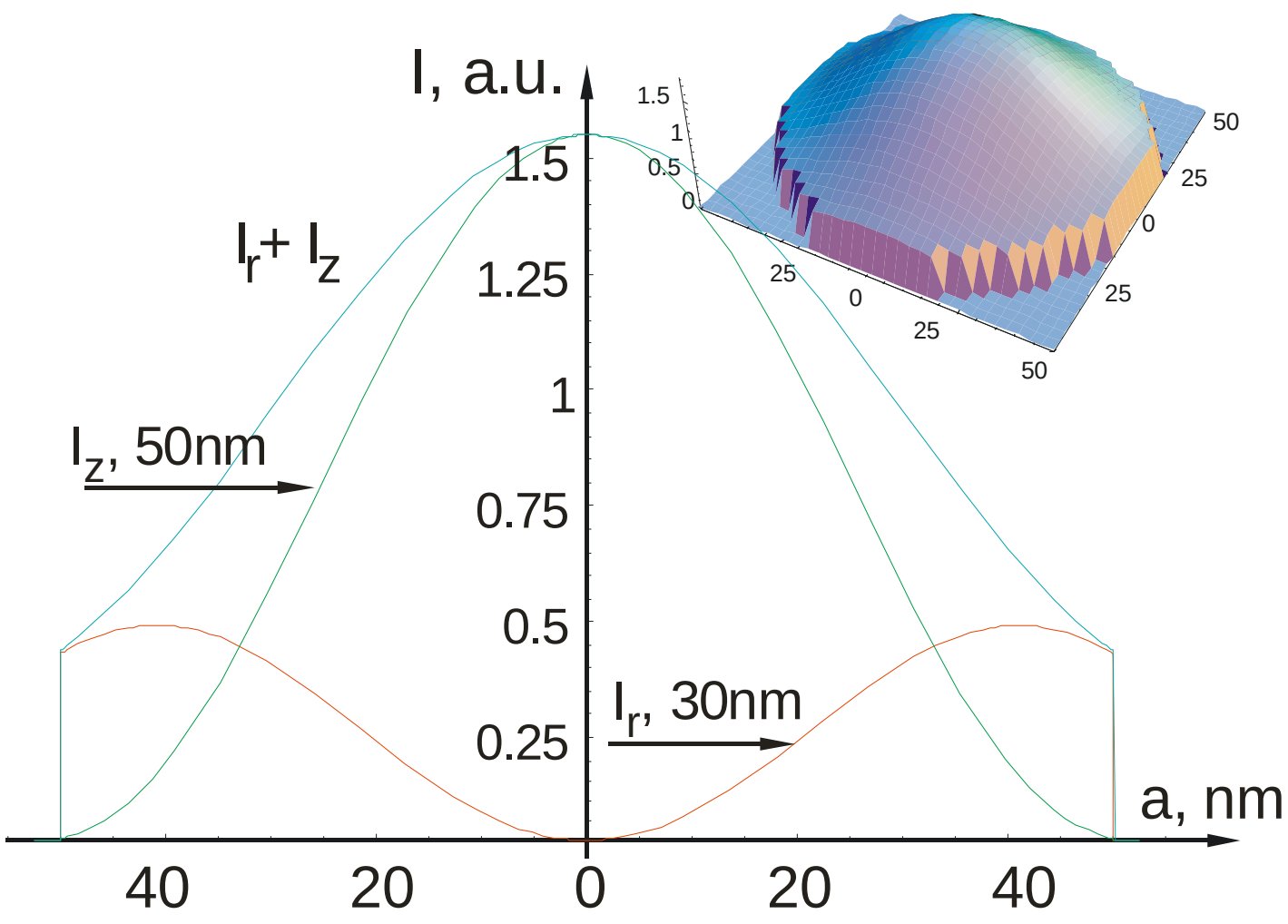

Fig. 4 Light field spatial structure inside the probe output aperture with angle α= $25^0$.

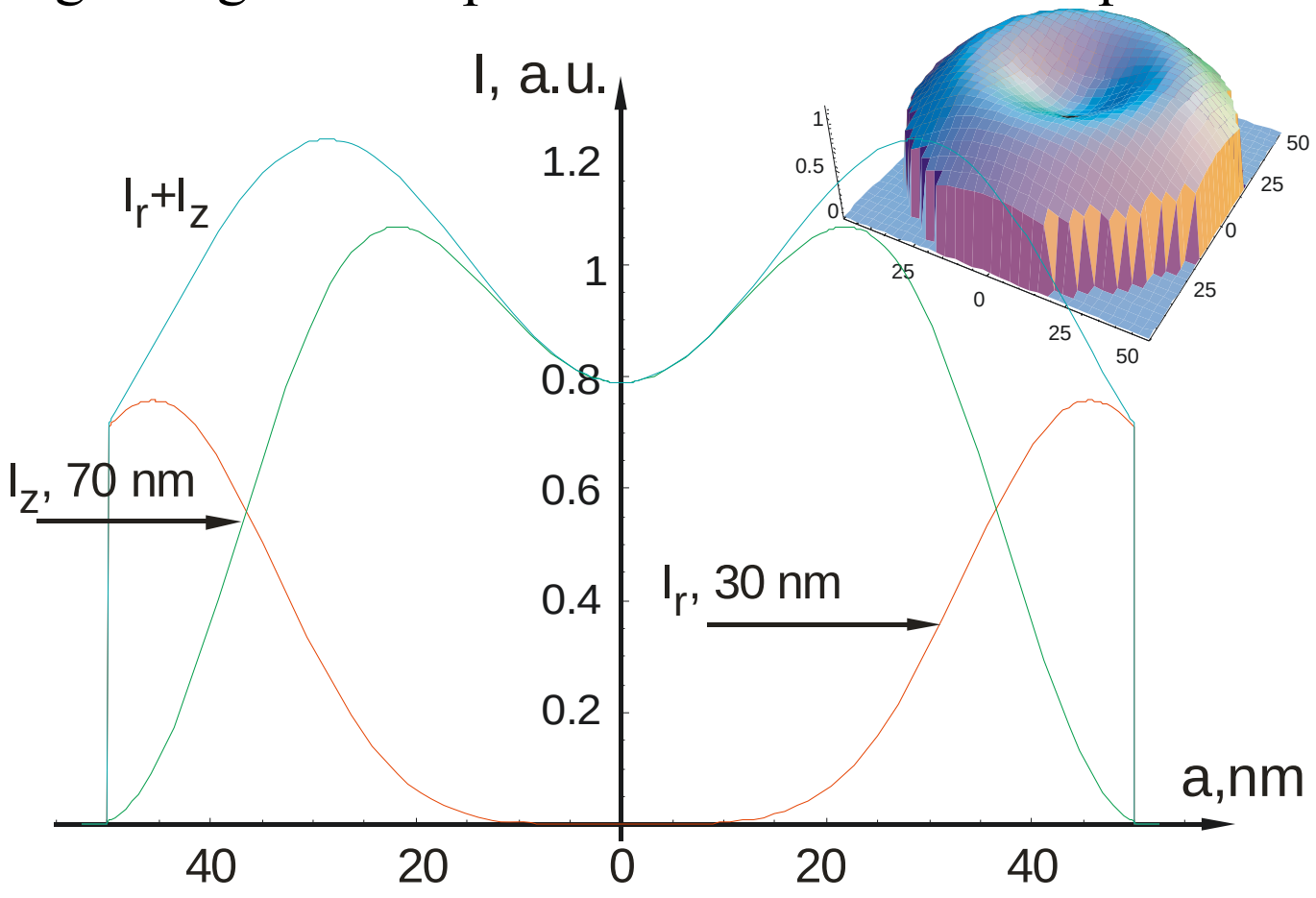

Fig. 5 Light field spatial structure inside the probe output aperture with angle α= $55^0$.

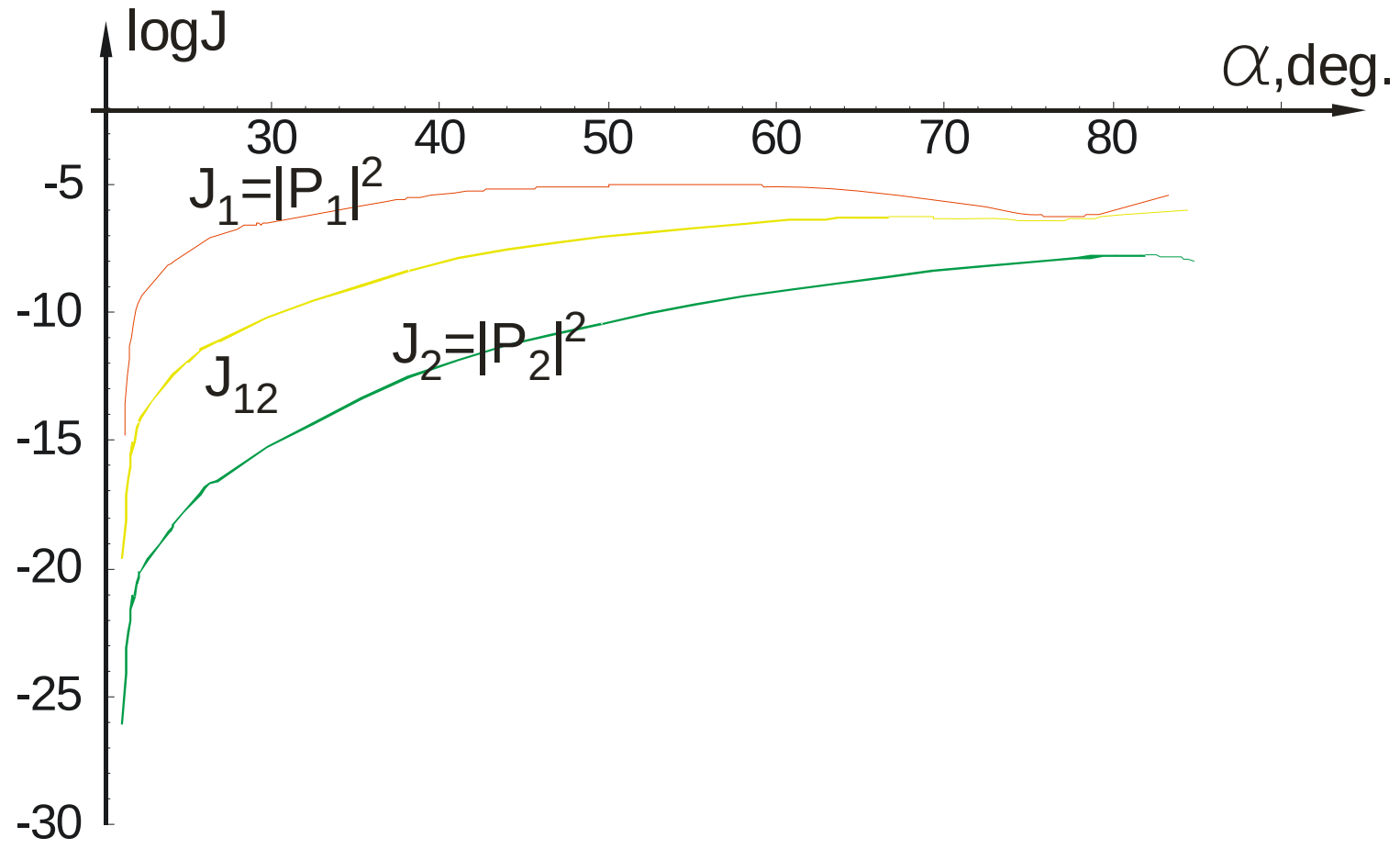

Fig. 6 Field amplitudes in the Eq. (9), Eq. (10). $|P_1(z)|^2$ - red line, $|P_2(z)|^2$ - green line, $2|P_1(z)P_2(z)\cos(\theta_1-\theta_2)|$ - yellow line.

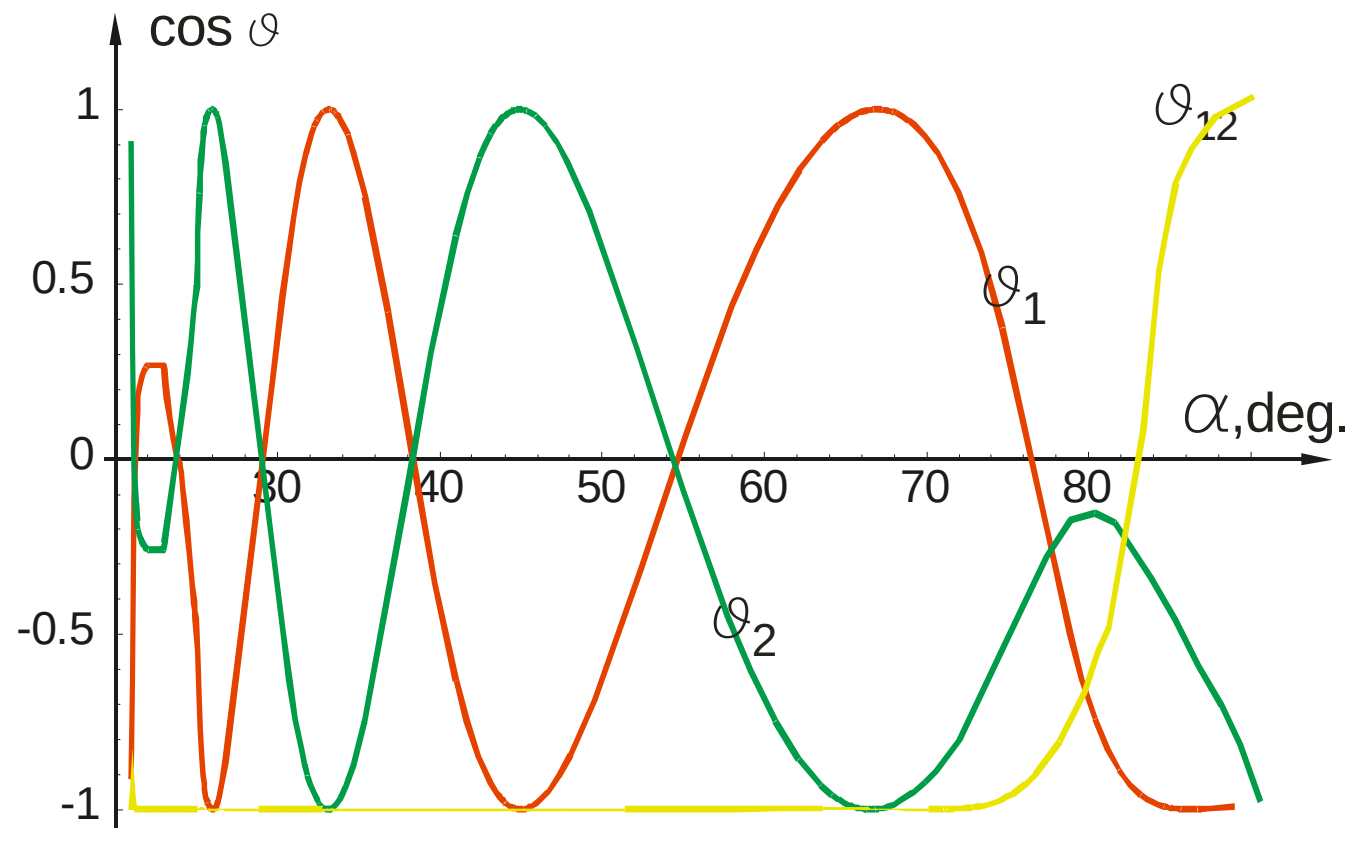

Fig. 7 Mode phases and their differences depending on the inclination α. It is seen that the phase relation between the two lowest light modes is beneficial for spatial compression at the inclination angle $\alpha \approx 75^{o}$, where the mode amplitudes became comparable on magnitude.

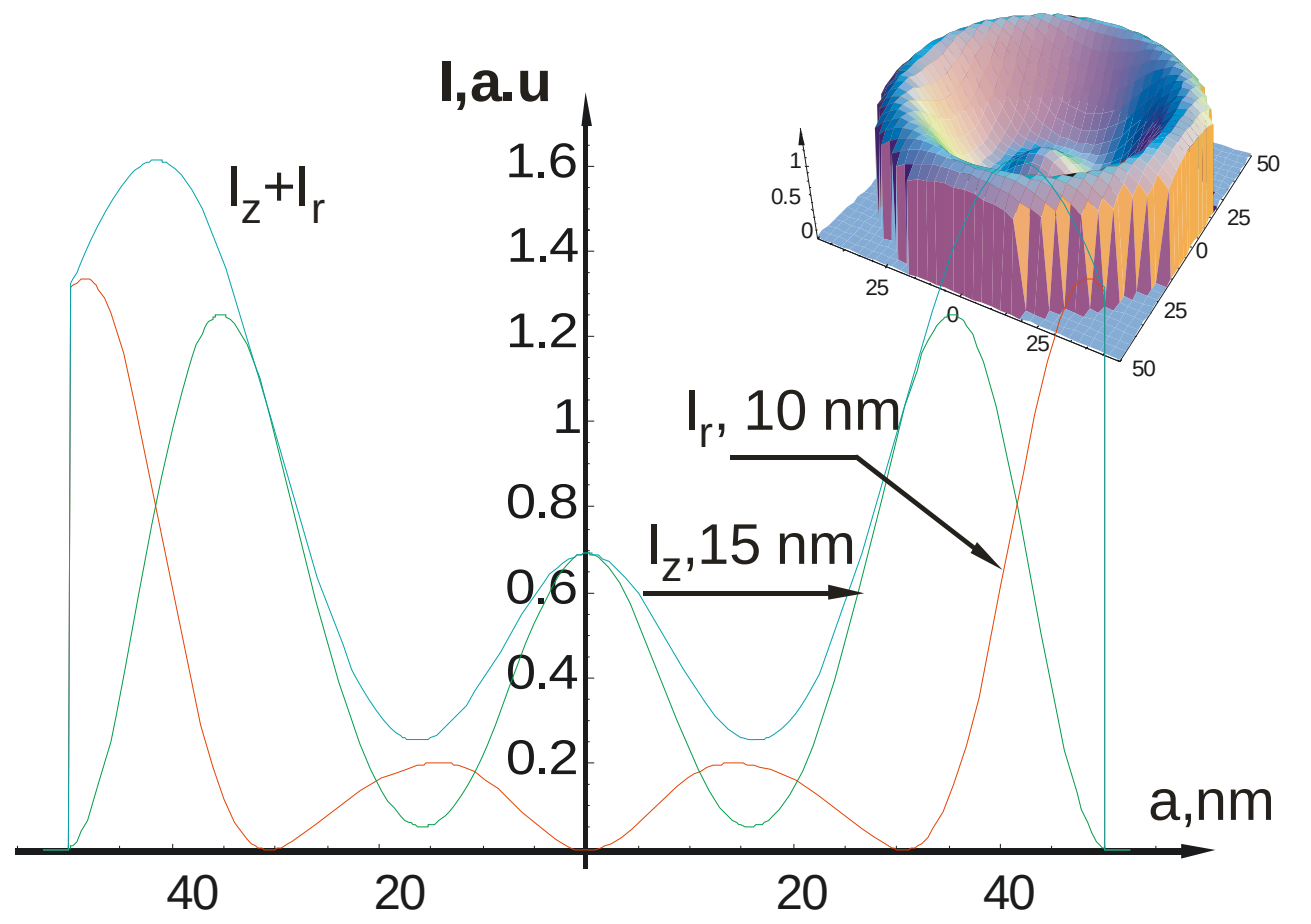

Fig. 8 Light field spatial structure inside the output aperture of probe with inclination angle $\alpha = 75^{0}$.

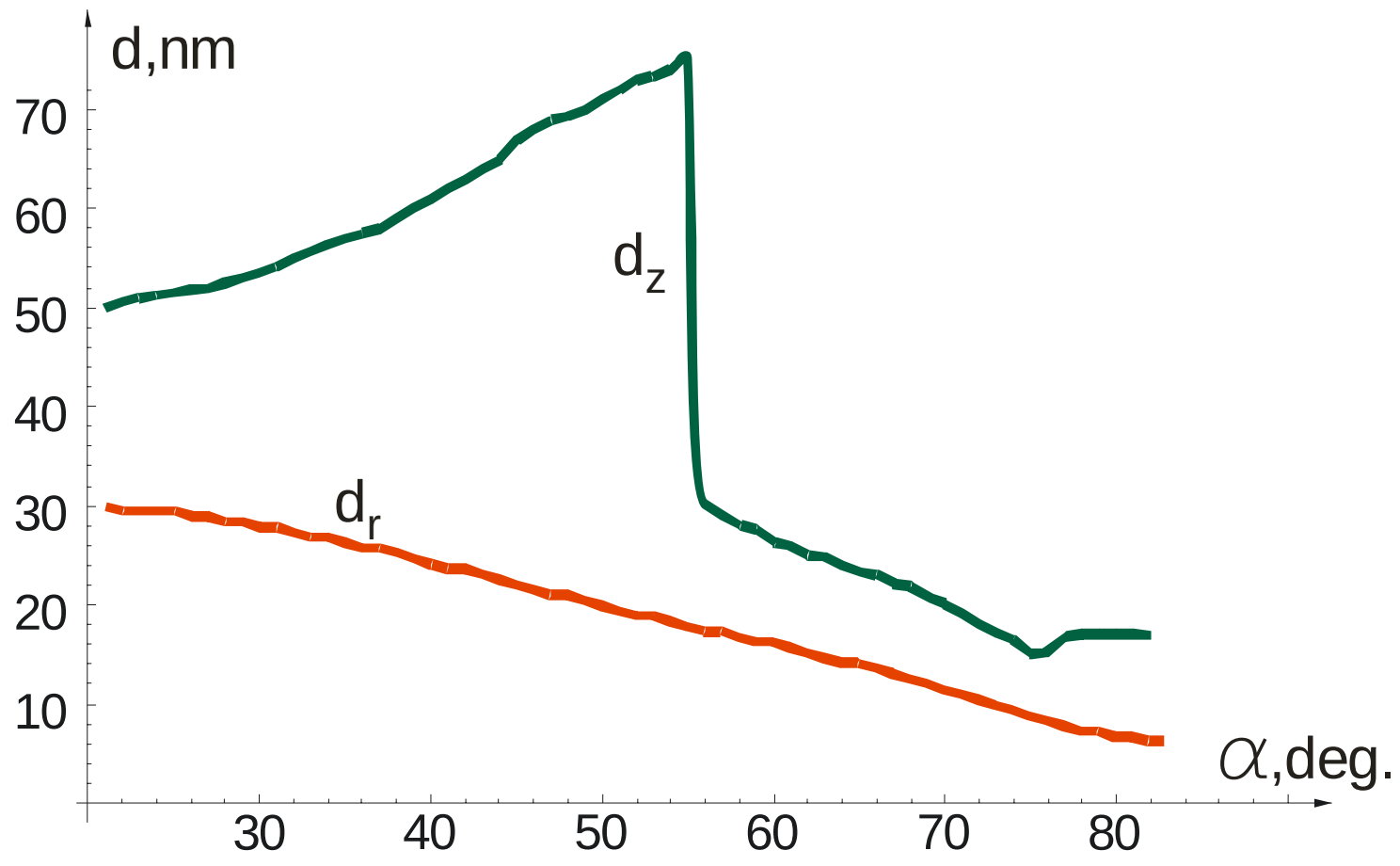

Fig.9. Spatial width of $d_z(L)$ (green) and $d_r(L)$ (red). There are taken into account Raleigh criteria of line resolution. There is a minimum spatial width of longitudinal field intensity at the $\alpha \approx 75^0$.